\documentclass{article}

\usepackage{arxiv}

\usepackage[utf8]{inputenc} 
\usepackage[T1]{fontenc}    
\usepackage{hyperref}       
\usepackage{url}            
\usepackage{booktabs}       
\usepackage{amsfonts}       
\usepackage{nicefrac}       
\usepackage{microtype}      
\usepackage{color, colortbl}
\usepackage[table]{xcolor}
\definecolor{Gray}{gray}{1.0}
\usepackage{float}
\usepackage{tabu}
\usepackage{graphicx}
\usepackage{algorithm}
\usepackage{cite}
\usepackage{amsmath,amssymb,amsfonts}
\usepackage{algorithmic}
\usepackage{textcomp}
\usepackage{booktabs, multirow}

\newcolumntype{L}{>{\centering\arraybackslash}m{0.08\linewidth}}
\newcolumntype{M}{>{\centering\arraybackslash}m{0.13\linewidth}}
\newcolumntype{N}{>{\centering\arraybackslash}m{0.09\linewidth}}
\newcolumntype{D}{>{\centering\arraybackslash}m{0.04\linewidth}}

\title{Consumer Wearables and Affective Computing for Wellbeing Support}

\author{
 Stanisław Saganowski, Przemysław Kazienko, Maciej Dzieżyc\\
 \textbf{Patrycja Jakimów, Joanna Komoszyńska, Weronika Michalska}\\
 Department of Computational Intelligence\\
 Faculty of Computer Science and Management\\
 Wrocław University of Science and Technology\\
 Wrocław, Poland \\
 \texttt{stanislaw.saganowski@pwr.edu.pl} \\
 \And
 Anna Dutkowiak, Adam Polak \\
 Faculty of Electronics\\
 Wrocław University of Science and Technology\\
 Wrocław, Poland \\
 \And
 Adam Dziadek, Michał Ujma \\
 Capgemini\\
 Wrocław, Poland \\
}

\begin{document}
\maketitle

\begin{abstract}
Wearables equipped with pervasive sensors enable us to monitor physiological and behavioral signals in our everyday life. We propose the WellAff system able to recognize affective states for wellbeing support. It also includes health care scenarios, in particular patients with chronic kidney disease (CKD) suffering from bipolar disorders. For the need of a large-scale field study, we revised over 50 off-the-shelf devices in terms of usefulness for emotion, stress, meditation, sleep, and physical activity recognition and analysis. Their usability directly comes from the types of sensors they possess as well as the quality and availability of raw signals. We found there is no versatile device suitable for all purposes. Using Empatica E4 and Samsung Galaxy Watch, we have recorded physiological signals from 11 participants over many weeks. The gathered data enabled us to train a classifier that accurately recognizes strong affective states.
\end{abstract}


\section{Introduction}
Each iteration of modern smart watches, wristbands, armbands, fitbands, headbands, chest straps, and patches is more powerful, precise, comfortable, and useful. At the same time, wearables are more affordable and easily available, ultimately becoming pervasive. Therefore, they may be applied in more and more domains. Their portability and low costs enable us to perform not only research in the lab but also large-scale field studies. 

Consequently, consumer wearables connected with smartphones are a perfect instrument for various wellbeing-related solutions. In particular, they may monitor our affective states and accordingly support decision making like environment adaptation (calming lighting if you are over-excited), alerting parents when a distant child is experiencing difficult times, warning very overweight people not to eat when they are having strong emotions or reporting to nephrologists about frequent changes in mood in patients with chronic kidney disease (CKD) or after kidney transplantation in order to adjust treatment or drug dosing.

To enable the aforementioned applications, we propose the WellAff system, Section \ref{sec:wellaff}. Its primary purpose is to monitor and recognize affective states like emotions as well as their changes in everyday life based on physiological sensor signals provided by off-the-shelf devices. 

For that purpose, we revised consumer wearables. The existing literature has focused on usefulness of commercially available devices in one  \cite{cajita2020feasible} or two domains \cite{peake2018critical}, or on validity of wearable sensors \cite{cosoli2020wrist}. In this work, however, we evaluate over 50 off-the-shelf wearable devices in terms of their usefulness and applicability in emotion and stress recognition, as well as sleep, meditation, and activity monitoring, Section \ref{sec:wearables}. We focus on the sensors built into the device, the signals they provide, and the availability of the raw data recorded or extracted with these sensors. This survey enabled us to select the most suitable devices for our further studies within the WellAff system: auxiliary Samsung Galaxy Watch and Empatica E4, as well as the target - Samsung Galaxy Watch 3.

An essential component of every machine learning-based recognition system is annotated data. There are only a few studies about emotion recognition in the field \cite{saganowski2020PerCom}. Therefore, we decided to perform a large-scale real life study. It was split into three phases, Fig. \ref{fig:system}: (1) data accumulation and a reasoning system about strong affective states, (2) recognition of these strong states and their more detailed self-annotations, and (3) final application of complex states recognition and change detection for wellbeing support.

One of the main challenges to overcome was: how to gather data about affect experienced in real life without a significant impact of the questioning procedure on experiencing itself? Hence, we believe that the annotation of affective states in real life should be as little intrusive as possible. Therefore, we apply in the WellAff system the concept of Ecological Momentary Assessment (EMA) \cite{shiffman2008ema}, which is in line with Descriptive Experience Sampling (DES). DES captures the inner experience of subjects in their natural environments using a random beeper invoking rare subject surveying \cite{hurlburt2002des}. We implemented this by means of randomness and decreased questioning frequency. EMA, in turn, is a data collection method in real-time and in real-world settings to avoid retrospective biases and gather ecologically valid data. EMA is particularly suited for studying substance use -- in our case, strong affect events.

\begin{table*}[ht]
\setlength{\tabcolsep}{4.2pt} 
\renewcommand{\arraystretch}{0.9} 
\centering
  \caption{The usefulness of consumer wearables in emotion (Emo), stress (Str), meditation (Md), sleep (Slp), and physical activity (Act) analysis. (Phy. raw sign.) denotes the availability of raw physiological signals; (*) marks wearables tested by us; (**) denotes the ECG sensor unlocked only in the South Korean market so far. Factors considered in grading: richness, sampling, and availability of relevant data, domain-related convenience, battery life.
  }
  \label{tab:wearables}
\scalebox{0.89}{
\begin{tabular}{
>{\centering\arraybackslash}m{2.9cm}|>{\centering\arraybackslash}m{1.5cm} >{\centering\arraybackslash}m{0.8cm} >{\centering\arraybackslash}m{3.3cm} >{\centering\arraybackslash}m{2.1cm} >{\centering\arraybackslash}m{3.5cm}|>{\centering\arraybackslash}m{0.4cm} >{\centering\arraybackslash}m{0.4cm} >{\centering\arraybackslash}m{0.4cm} >{\centering\arraybackslash}m{0.4cm} >{\centering\arraybackslash}m{0.4cm}
}
\hline
\textbf{Device*} & \textbf{Type} & \textbf{Release} & \textbf{Sensors} & \textbf{Phy. raw sign.} & \textbf{Other data} & \textbf{Emo} & \textbf{Str} & \textbf{Md} & \textbf{Slp} & \textbf{Act} \\ \midrule
Samsung Galaxy Watch 3*& Smart watch & 2020.08 & PPG, ECG** ACC, GYRO, BAR, AL, MIC, GPS & BVP & HR, ACC, GYRO, BAR, AL, MIC, GPS, STP & ++ & ++ & + & ++ & \cellcolor{gray!15}{+++} \\
\hline
Apple Watch 5*& Smart watch & 2019.09 & PPG, ECG, ACC, GYRO, BAR, MIC, GPS & - & HR, ACC, GYRO, BAR, MIC, GPS, STP, CAL & +  & ++  & + & ++  &  \cellcolor{gray!15}{+++} \\
\hline
Fossil Gen 5*& Smart watch & 2019.08 & PPG, ACC, GYRO, ALT, AL, MIC, GPS & BVP & HR, ACC, GYRO, ALT, AL, MIC, GPS, STP & ++ & ++ & + & ++ & \cellcolor{gray!15}{+++} \\
\hline
Garmin Fenix 6X Pro& Smart watch & 2019.08 &  PPG, SpO2, ACC, GYRO, ALT, AL, GPS & BVP, SpO2 & HR, ACC, GYRO, ALT, AL, GPS, STP & ++ & ++ & + & ++ & \cellcolor{gray!15}{+++} \\
\hline
Samsung Galaxy Watch*& Smart watch & 2019.08 & PPG, ACC, GYRO, BAR, AL, MIC, GPS & BVP & HR, ACC, GYRO, BAR, AL, MIC, GPS, STP & ++ & ++ & + & ++ & \cellcolor{gray!15}{+++} \\
\hline
Polar OH1 & Armband & 2019.03 & PPG, ACC & BVP & PPI, ACC & ++ & ++ & + & ++ & ++ \\
\hline
Samsung Galaxy Fit E* & Fitband & 2019.02 & PPG, ACC & - & HR & - & + & + & + & + \\
\hline
Garmin HRM-DUAL & Chest strap & 2019.01 & ECG & ECG & RRI & ++ & ++ & + & + & + \\
\hline
Muse 2*& EEG headband & 2019.01 & EEG, PPG, SpO2, ACC, GYRO & EEG, BVP, SpO2, ACC, GYRO & HR & ++ & ++ & \cellcolor{gray!15}{+++} & + & - \\
\hline
Fitbit Charge 3* & Fitband & 2018.10 & PPG, ACC, GYRO, ALT & - & HR, ACC, ALT & + & ++ & + & ++ & ++ \\
\hline
Garmin VivoActive 3 Music* & Smart watch & 2018.06 & PPG, ACC, GYRO, BAR, GPS & - & HR, PPI, RSP, ACC, STP, CAL& + & ++ & + & ++ & ++ \\
\hline
Oura ring* & Smart ring & 2018.04 & PPG, ACC, GYRO, TERM & - & HR, PPI, SKT, SP& - & ++ & + & \cellcolor{gray!15}{+++} & + \\
\hline
Moodmetric* & Smart ring & 2017.12 & EDA, ACC & EDA & STP& + & ++ & - & + & - \\
\hline
DREEM& EEG headband & 2017.06 & EEG, PPG, SpO2, ACC & EEG, BVP, SpO2, ACC & HR & ++ & ++ & \cellcolor{gray!15}{+++} & ++ & - \\
\hline
Polar H10 & Chest strap & 2017.03 & ECG, ACC & ECG & RRI, ACC & ++ & ++ & + & ++ & ++ \\
\hline
VitalPatch & Chest patch & 2016.03 & ECG, ACC, TERM & ECG, SKT & HR, RRI, EDR, STP & ++ & ++ & + & ++ & + \\
\hline
Sony SmartBand 2 & Fitband & 2015.09 & PPG, ACC & BVP & HR, PPI, ACC & + & ++ & + & ++ & ++ \\
\hline
Empatica E4* & Wristband & 2015 & PPG, EDA, ACC, TERM & BVP, EDA, SKT &  HR, PPI, ACC, tags & \cellcolor{gray!15}{+++} & \cellcolor{gray!15}{+++} & + & \cellcolor{gray!15}{+++} & ++ \\
\hline
Microsoft Band 2 & Smartband & 2014.10& PPG, EDA, ACC, GYRO, TERM, BAR, ALT, AL, UV & BVP, EDA, SKT & HR, PPI, ACC, GYRO, BAR, ALT, AL, STP, CAL, UV & \cellcolor{gray!15}{+++} & \cellcolor{gray!15}{+++} & + & \cellcolor{gray!15}{+++} & ++ \\
\hline
Samsung Gear Live & Fitband & 2014.06 & PPG, ACC, GYRO & BVP & HR, ACC, GYRO, STP & ++ & ++ & + & ++ & ++ \\
\hline
Philips DTI-2 & Wristband & 2014.03 & EDA, ACC, TERM, AL, AT & EDA & ACC, TEMP, AL, AT & + & ++ & - & ++ & + \\ \bottomrule
\end{tabular}}
\end{table*}

\section{Signals in Emotion, Meditation, Stress, Sleep, and Activity Analysis}
\label{sec:signals}

Wearables are equipped with sensors that provide physiological, behavioral, and environmental data, Tab. \ref{tab:wearables}. These sensors are EEG - electroencephalogram, PPG - photoplethysmograph\footnote{Some producers like Samsung in their Galaxy Watch series provide the raw PPG signal in the form of light intensity stream.} delivering BVP - Blood Volume Pulse signal and derived signal PPI - peak-to-peak intervals (a.k.a. HRV - heart rate variability, or IBI - interbeat interval), ECG - electrocardiograph providing RRI - R-R intervals, EDA - electrodermal activity sensor (a.k.a GSR - galvanic skin response), SpO2 - blood oxygen saturation sensor, ACC - accelerometer, GYRO - gyroscope, TERM - thermometer providing SKT - skin temperature, BAR - barometer, ALT - altimeter, AL - ambient light sensor, AT - ambient temperature sensor, MIC - microphone, MAG - magnetometer, UV - ultraviolet, and GPS. They can also provide other data derived from the monitored signals: HR - heart rate (extracted either from BVP/PPI or ECG/RRI), STP - number of steps, RSP - respiration rate, EDR - RSP from ECG, CAL - calories burned, SP - sleep phases.

\textbf{Emotions}.
The EEG outperforms other signals in terms of usefulness for emotion recognition \cite{nakisa2018long, soroush2017review}. However, recently, studies tend to employ wearables \cite{saganowski2020PerCom}, which provide various bio-signals and additional environmental data. Most often, ECG/BVP and EDA signals are utilized \cite{nalepa2019analysis, setiawan2018framework, fernandez2019emotion}. Those signals can be supplemented with ACC, GYRO SKT, RSP \cite{schmidt2018labelling, he2017emotion, maier2019deepflow, schmidt2018introducing, albraikan2018iaware}, as well as with UV, GPS, and MIC data \cite{kanjo2019deep}.

\textbf{Stress}.
Bio-signals related to stress include EEG, ECG, BVP, EDA, SKT, and RSP \cite{giannakakis2019review, can2019stress}. The best stress detection accuracy in the field can be achieved when using ECG/BVP and EDA signals together \cite{ghosh2015annotation, golgouneh2019fabrication, can2019continuous}. Albeit, using the ECG/BVP or EDA signal solely also provides satisfactory results \cite{zangroniz2018estimation, zangroniz2017electrodermal}.

\textbf{Meditation}.
The most appropriate physiological signal to study meditation is EEG \cite{sharma2019eeg, hagad2019deep, harne2019svm}. Other useful signals are ECG \cite{leonard2019changes, alawieh2019real}, BVP \cite{yu2018unwind}, and EDA \cite{pavani2019study}.

\textbf{Sleep}.
There are two main methods for gathering data in sleep studies: polysomnography (PSG) and actigraphy (ACG) \cite{grandner2019actigraphic}. A gold standard is PSG, which uses advanced medical equipment to measure EEG, EOG, EMG, ECG. However, due to the complexity, size, wiring, and cost of the equipment, PSG cannot be applied outside the lab. For field studies, ACG is an appropriate measurement method based on the movement signals, i.e., ACC data. More recent studies also utilize BVP \cite{assaf2018sleep, de2019sleep}, SKT \cite{wei2019you}, and MIC \cite{castaldo2019investigating, camci2019abnormal}. Data from SpO2, ECG, RSP, and MIC allow us to diagnose the sleep apnea \cite{mendonca2018review, nikkonen2019artificial, erdenebayar2019deep}.

\textbf{Physical activity} research commonly focuses on the following problems: identification, tracking, and quantification. The  most frequently used sensors, providing data to tackle these problems are ACC, GYRO, and GPS \cite{Allahbakhhi2020Accelerometer, aroganam2019review, Murakami2019AccuracyO1, s140610146}, as well as PPG \cite{WearablePham2020, Mller2019HeartRM, ARUNKUMAR2020101790}, and ECG \cite{SelenaAccuracycommerciallyheartrate2019}.

\section{Consumer Wearables for Activity and Affective Computing}
\label{sec:wearables}

When considering whether a device is appropriate for lab or field studies related to a given problem, many factors have to be taken into account. The most important is the ability to provide relevant and rich data in a raw format. Many producers develop dedicated applications for end-users to allow them convenient monitoring of the data collected by various sensors. However, this data is usually processed in some way (e.g., aggregated, filtered, smoothed) and very rarely available to download. Thus, not useful for the research purpose. Some manufacturers provide the ability to access raw data by integrating with a device over Bluetooth or another protocol, e.g., Empatica E4, Garmin HRM-DUAL, Moodmetric. Such integration is sufficient to gather data for the research but still has some limitations, e.g., the sensors' sampling frequency cannot be adjusted. Only a few producers provide an SDK (software development kit) that allows us to develop our own software and embed it into devices, e.g., Samsung's smart watches running Tizen operating system (OS), and Wear OS devices. See Tab.~\ref{tab:wearables}, column \textit{Phy. raw sign.} for the list of raw physiological signals provided by particular consumer devices. Overall, the more signals/sensors related to a considered problem, the better.

Tab. \ref{tab:wearables} only contains devices that are \textit{portable} and grant access to the gathered data. Of course, the \textit{portability} argument is subjective. Ten years ago, a backpack with sensors was considered portable. Today, portables have to be small and handy. Perhaps, in five years, portable would mean barely visible or even implanted.

Raw signals, especially sampled with high frequency, provide more flexibility in research; hence, their availability is essential. Portability, in turn, is crucial for field studies in particular long-term ones. Some devices provide signals in real-time, whereas the others after the session termination. All wearables in Tab. \ref{tab:wearables}, except DTI-2, are equipped with Bluetooth Low Energy (BLE); few also provide Wi-Fi, LTE, and NFC connectivity. 
We examined devices marked with * in terms of raw signals availability. For other devices, we relied on the official producers' or other external information. 

To sum up, there is no off-the-shelf device equipped well enough for a proper analysis in all domains, i.e., emotion, stress, meditation, sleep, and physical activity. Producers design them for specific market needs and use sensors and electronic units of different quality and often provide access only to selected data. 
Moreover, the available signals may have too low sampling frequency. For example, the averaged over 5 mins PPI and HR in Oura Ring is insufficient for accurate emotion recognition (-) but is acceptable for approximate stress detection (++). Furthermore, with additional sleep-related data Oura is very good for sleep analysis (+++). The same refers to Apple Watch 5 that provides HR data every 5 secs (and no other physiological signals) --- too rarely for emotions but good enough for activity studies (+++). 

The best wearables for emotion, stress, and sleep research appear to be the relatively old Empatica E4 and Microsoft Band 2. The former will be replaced with EmbracePlus this year; the latter is no longer supported. EEG headbands, Muse 2, and DREEM are the best choice for meditation. Although DREEM is intended for sleep studies, we argue that it is not very comfortable. Oura Ring is more suitable for undisturbed sleep analysis. The smart watches by Apple, Fossil, Garmin, and Samsung have proper sensors and provide the best data for activity investigations. 

We have also considered some other devices, like Emotiv Epoc+*, NeuroSky, emWave2, Honor Band 4*, Xiaomi Mi Band 3*, Xiaomi Mi Band 5*, Polar A370*, Fitbit Blaze*, and 20 others. However, they do not offer access to the data or have other drawbacks. Therefore, they have not been included in Tab. \ref{tab:wearables}. 

In our studies focused on the affect recognition, we have decided to use a few Empatica E4 devices and dozens of Samsung Galaxy Watches.
Empatica E4 offers the greatest number of sensors among the devices available in the market but is also a few times more expensive than smart watches. On the other hand, Samsung Galaxy Watch lacks only the EDA and SKT sensors, but offers some other data like BAR or AL (eventually, should also offer ECG), and allows us to build custom watch applications.

\section{WellAff: a System for Wellbeing Support based on Affective Computing}
\label{sec:wellaff}
\begin{figure*}[ht]
  \centering
    \includegraphics[width=0.7\textwidth]{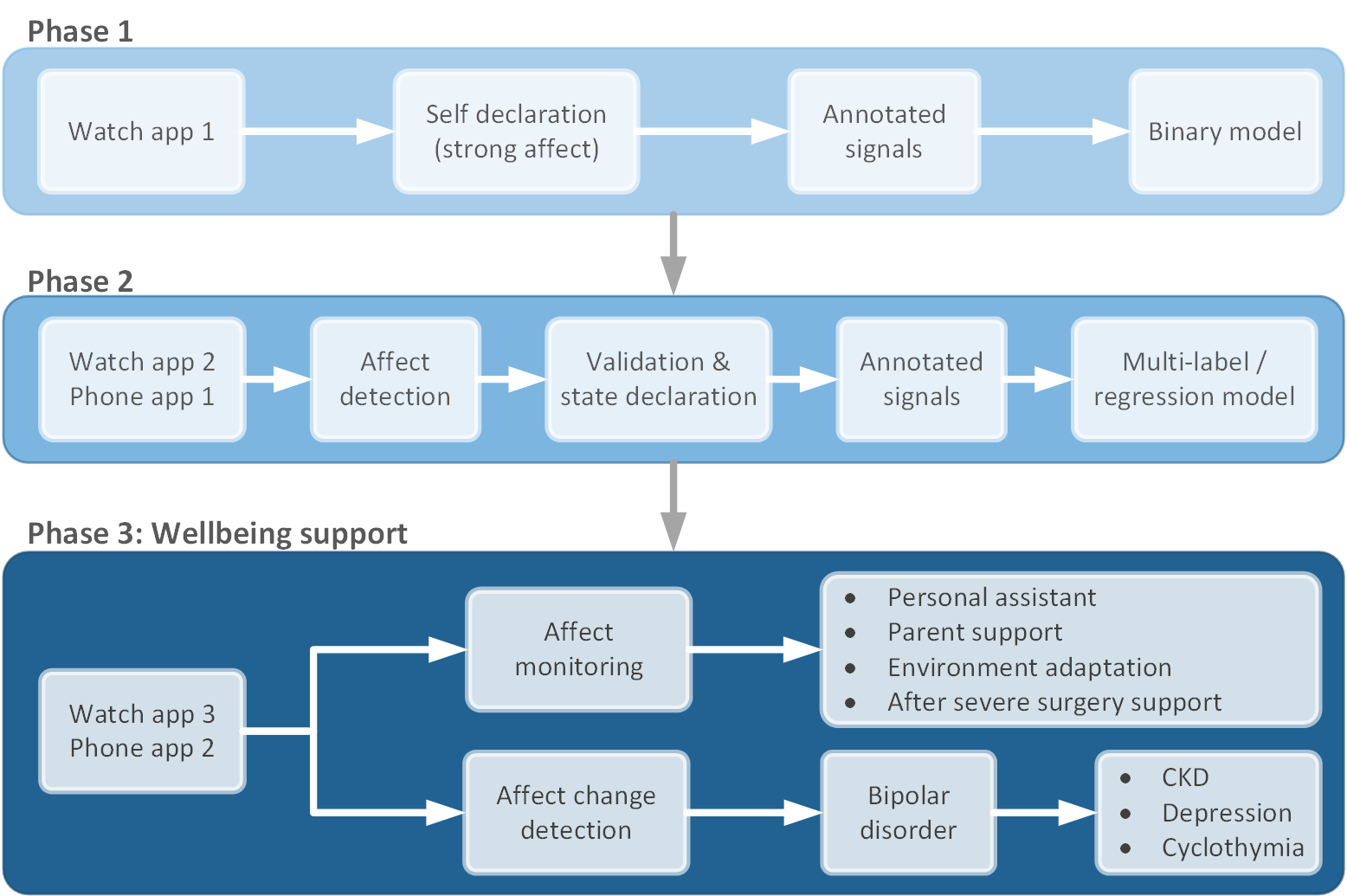}
    \caption{A WellAff system for wellbeing support through monitoring and recognition of affective states in everyday life.}
    \label{fig:system}
\end{figure*}

The general idea of the WellAff system, including its development phases, is depicted in Fig. \ref{fig:system}.
The goal of the first phase is to train a classification model able to recognize strong affective states experienced in everyday life based on physiological signals from wearables (smart watches, wristbands). We have developed a watch application for Samsung Galaxy watches that allows the participants to mark both strong affective states and neutral ones (low arousal). An analogous, built-in functionality is provided by Empatica E4 by means of the dedicated button. Having physiological signals annotated with very intensive and neutral states, we can build a binary classifier recognizing whether a participant is currently experiencing a strong affect or not.

The trained classifier is directly used as a predictor in the second phase. It is a crucial part of the integrated smart watch and smartphone application. It continuously monitors physiological and behavioral signals. If, based on them, it recognizes a strong affect, a notification appears on the watch with a request to confirm the prediction. The participants may either reject such a suggestion (then we mark a false positive prediction) or accept it. If they agree, the phone application provides the EMA-type questionnaire \cite{shiffman2008ema} - a list of affective states along with the open text option. Additionally, to make the system less intrusive and avoid too frequent surveying, we apply some randomness and minimal idle time between consecutive questioning. In this way, we are able to identify moments of experiencing strong affects in real life without much interfering. As a result, we obtain a new data set consisting of signals and new, more comprehensive manual annotations. Some affective states like emotions can often co-occur. Having this in mind and based on the data collected, we build various prediction models with distinct outputs: binary, multi-class, multi-label, regression, multi-dimensional, structured output, e.g., a sequence of states/values. The model selection depends on the use-case (wellbeing support) that we want to apply in the final phase of the WellAff system.

The learned models are used in the third phase as a component of another wearable and mobile application. Similarly to the previous stage, the system is able to analyze signals provided by wearable sensors and to provide more complex predictions: different co-occurring affective states, their levels, and transitions -- meaningful changes. These real-time predictions are used in various applications related to wellbeing, personal, and health support. For example, bipolar disorders and related emotional distress may result from brain metabolic disturbances caused by uremic toxins accumulation. This, in turn, can be a symptom of Chronic Kidney Disease (CKD) or problems with treatment after renal transplantation. Being able to recognize the mixed affective states at an early stage, we could suggest CKD patients to go on dialysis, thus, avoiding significant toxins accumulation and consequently improving their mood and quality of life.

So far, we have completed the first phase with preliminary results, see Section \ref{sec:results}.

\section{Affect Recognition in Everyday Life - Experimental Results}
\label{sec:results}

To implement the entire WellAff idea, we started the first phase in spring 2020. For the purpose of all our WellAff-related studies, we received bioethics committee approval no. 149/2020 from Wroclaw Medical University.

\subsection{Learning Dataset Collection}

In the first phase, we have used Empatica E4 and Samsung Galaxy Watch 2019 wearables. The Empatica is among the most appropriate devices for affect recognition, but not very convenient for the end-user, while Samsung Galaxy Watch is user-friendly and lacks only EDA and SKT signals, see Tab.~\ref{tab:wearables}. The participants used a dedicated smart watch application to mark with timestamps the affective states they experience in their life. Additionally, they filled up an online questionnaire to report the type of the state (especially strong affect vs. neutral) and a possible delay between the experiencing affect and marked timestamp. If the delay exceeds 60 seconds, the events are excluded from the study, because in our opinion, it is hard to accurately estimate the delay after a couple of minutes from the strong affective event.

\begin{figure*}[ht]
  \centering
    \includegraphics[width=0.9\textwidth]{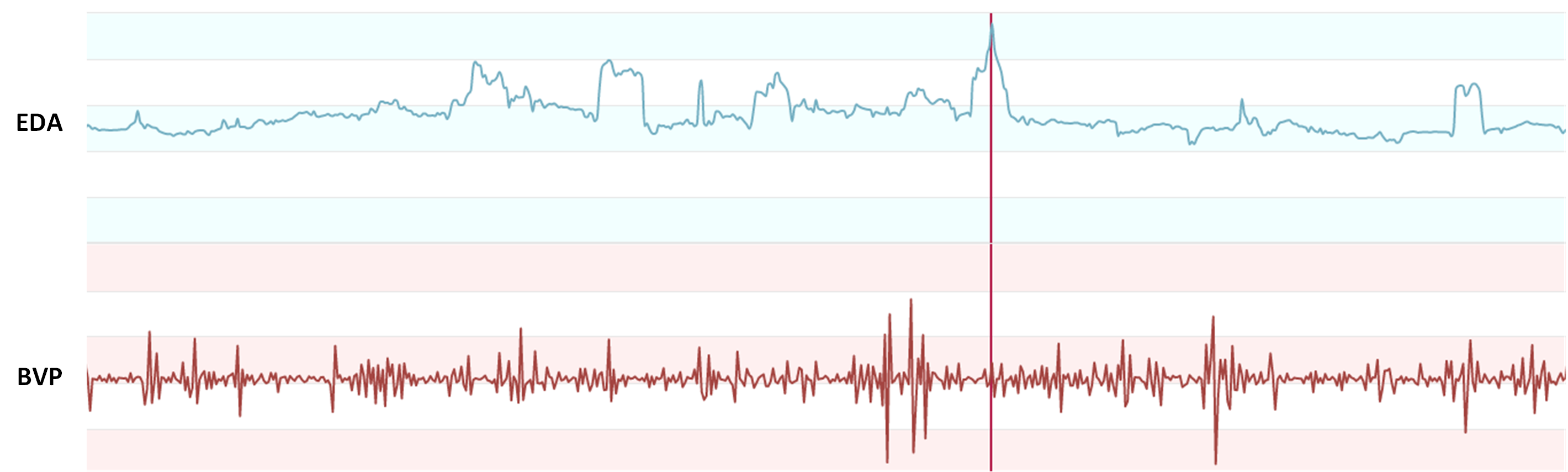}
    \caption{An example of strong affect (fear for the choked child) experienced in real life and marked by the participant. The body response in term of peak of the EDA signal can be easily observed.}
    \label{fig:example}
\end{figure*}

At the moment, we have gathered 294 affective samples (206 - strong affect, 75 - neutral or calm state, 13 samples were marked by mistake, therefore excluded) from six female and five male subjects, aged 23-53. During the three-month span, we have recorded over 3000 hours for each physiological and behavioral signal. Finally, we used signals from the following sensors: PPG (light intensity for Samsung Galaxy Watch, BVP for Empatica), ACC, TERM, and EDA (Empatica only), AL, GYRO, and BAR (Samsung only). An example of physiological signals during a strong affect -- fear for the choked child is depicted in Fig. \ref{fig:example}. The participant's body response in the form of a peak in the EDA signal can be easily observed. The data collection in this phase is still on-going, and new samples are gathered every day.

\subsection{Model for Strong Affect Recognition}

From all the recordings, the 60-second signal windows preceding the reported affective states have been extracted. Then, the signals were pre-processed using winsorization, Butterworth low-pass filter with 10Hz cut-off, and min-max normalization. The stratified sampling has been applied to divide 281 samples into training and test sets with 80:20 ratio and 5-fold cross-validation. The cases in the training set were balanced using the SMOTE \cite{chawla2002smote} method. Next, over 60 features were extracted from BVP, EDA, SKT, and ACC signals using BioSPPy\footnote{https://biosppy.readthedocs.io}, pyphysio\footnote{https://github.com/MPBA/pyphysio}, and SciPy\footnote{https://www.scipy.org} libraries. After feature selection and reduction, the binary classification (strong affect vs. neutral state) has been performed. The initial calculations performed with the tree-based models wrapped with the AdaBoost ensemble algorithm provided 91\% F1-measure efficiency.

\section{Conclusions}
Consumer wearables are yet to match the medical-level devices in terms of sensor and signal quality.
The main advantage of wearables is their portability, ubiquity, and multiple sensors enabling large-scale multimodal studies in everyday life. There are suitable wearables for each of the research topics considered in this paper. The initial results suggest that we can benefit from the use of wearables in wellbeing support.

The nearest future work will cover: (i) additional sampling of neutral states to deliver a better class balance, (ii) further collection of new samples with new subjects and diverse conditions, (iii) training and validation of the end-to-end multimodal deep learning models as well as their simplification for reasoning purposes on smart watches and smartphones (some preliminary results have already been obtained), (iv) implementation of the second and third phase, (v) gathering data from CKD-suffering patients.

\subsection*{Acknowledgments}
This work was partially supported by the National Science Centre, Poland, project no. 2016/21/B/ST6/01463; and the statutory funds of the Dept. of Computational Intelligence, Wroclaw University of Science and Technology.

\bibliographystyle{unsrt}
\bibliography{emotions}

\end{document}